\pdfoutput=1

\documentclass[superscriptaddress,amsmath,amssymb,twocolumn]{revtex4-1}

\usepackage{graphicx}
\usepackage{chemarr}
\usepackage{times,longtable}
\usepackage{hyperref}
\usepackage{color}
\usepackage{afterpage}
\usepackage{soul,comment}
\usepackage[usenames,dvipsnames]{xcolor}
\usepackage{colortbl}
\usepackage{pbox}
\usepackage{capt-of}

\graphicspath{{large/}}

\bibpunct{[}{]}{,}{n}{}{,}

\newcommand{\avg}[1]{\left\langle #1 \right\rangle}
\renewcommand{\bar}[1]{\overline{#1}}
\newcommand{\eff}{{\rm eff}}

\renewcommand{\hat}[1]{\widehat{#1}}

\usepackage{chemarr}
\usepackage{epsf,mathtools}
\usepackage{graphicx}
\usepackage{dcolumn}
\usepackage{bm}

\begin{document}

\title{\textsf{Translating ceRNA susceptibilities into correlation functions}}

\author{Araks Martirosyan}
\affiliation{Dipartimento di Fisica, Sapienza Universit\`a di Roma, Rome, Italy}
\affiliation{VIB-KU Leuven Center for Brain \& Disease Research, Leuven, Belgium}

\author{Matteo Marsili}
\affiliation{The Abdus Salam International Centre for Theoretical Physics, Trieste, Italy}

\author{Andrea De Martino}
\affiliation{Soft and Living Matter Lab, Institute of Nanotechnology (CNR-NANOTEC), Consiglio Nazionale delle Ricerche, Rome, Italy}
\affiliation{Human Genetics Foundation, Turin, Italy}

\begin{abstract}
Competition to bind microRNAs induces an effective positive crosstalk between their targets, therefore known as `competing endogenous RNAs' or ceRNAs. While such an effect is known to play a significant role in specific conditions, estimating its strength from data and, experimentally, in physiological conditions appears to be far from simple. Here we show that the susceptibility of ceRNAs to different types of perturbations affecting their competitors (and hence their tendency to crosstalk) can be encoded in quantities as intuitive and as simple to measure as correlation functions. We confirm this scenario by extensive numerical simulations and validate it by re-analyzing PTEN's crosstalk pattern from TCGA breast cancer dataset. These results clarify the links between different quantities used to estimate the intensity of ceRNA crosstalk and provide new keys to analyze transcriptional datasets and effectively probe ceRNA networks {\it in silico}. 
\end{abstract}

\maketitle

\section*{\textsf{Introduction}}

\begin{figure*}
\begin{center}
\includegraphics[width=0.9\textwidth]{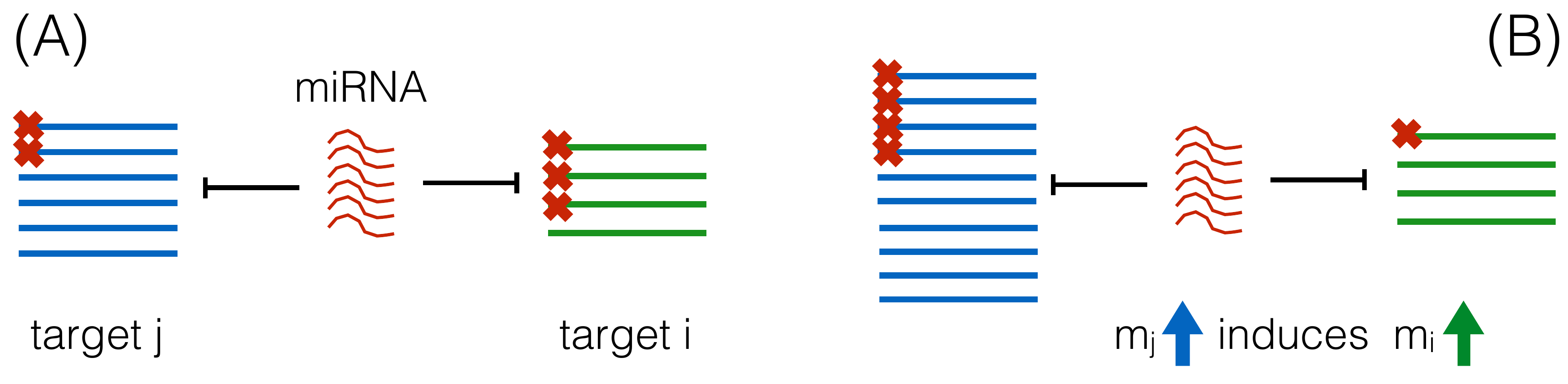}
\end{center}
\caption{(A) A miRNA species interacts with two different target RNAs who are competing to bind it, partially repressing their pools (repression being represented by red crosses). (B) In specific conditions, an increase in the level of one of the targets can induce a de-repression of the competitor, thereby establishing an effective positive coupling between the targets. Note that a similar effect can in principle be obtained between ceRNAs that are not co-regulated by the same  miRNA species through chains of miRNA-mediated interactions. \label{zero}}
\end{figure*}

MicroRNAs (miRNAs) are small non coding RNA (ncRNA) molecules that post-transcriptionally regulate a significant portion of the eukaryotic transcriptome via sequence-specific, protein-mediated binding in the cytoplasm \cite{flynt}. Their primary effects on coding transcripts consist in inhibiting translation and fostering degradation \cite{jonas}. Long ncRNAs, instead, can transiently sequester miRNAs, thereby altering their availability and overall repressive potential \cite{guil}. Following early observations concerning small regulatory RNAs in plants and bacteria \cite{franco,hwa}, competition to bind miRNAs has been hypothesized to cause an effective positive interaction (`crosstalk') between their coding and/or non-coding targets that may directly affect protein levels \cite{salmena} (see Fig. \ref{zero}A,B). Several experimental and modeling studies have clarified the conditions under which such a  scenario may become biologically relevant, highlighting specifically how molecular levels and kinetic heterogeneities may control it \cite{figl1,bosia,jens,bosson,vera,marti,marti2}. So far, such a `ceRNA effect' (whereby ceRNA stands for `competing endogenous RNA') has been quantitatively validated in cases of differentiation \cite{legnini}, disease \cite{tay} or in presence of unphysiologically large transcriptional inputs \cite{denzler}. Its significance in standard physiological conditions is therefore subject to scrutiny \cite{bartel2}. 

A major difficulty in detecting the ceRNA effect unambiguously in experiments or data lies in the fact that it should be disentangled from other mechanisms that may bear a similar impact, i.e. an effective positive coupling, on transcripts. Imagine a network of $N$ ceRNA species interacting with $M$ miRNA species. ceRNA levels $m_i$ ($i=1,\ldots, N$) fluctuate stochastically in time due to random synthesis and degradation events and to interactions with miRNAs, whose levels are also subject to random fluctuations. Denoting by $\avg{\cdot}$ the time-average in the steady state, an effective ceRNA-ceRNA dependence can be signaled by a statistical correlation coefficient such as Pearson's \cite{bosia}, i.e.
\begin{equation}\label{rho}
\rho_{ij}=
\frac{\avg{m_i m_j}-\avg{m_i}\avg{m_j}}{[(\avg{m_i^2}-\avg{m_i}^2)(\avg{m_j^2}-\avg{m_j}^2)]^{1/2}},
\end{equation}
with the idea that, if $\rho_{ij}$ is large enough, a perturbation altering the level of ceRNA $j$ will cause part of the miRNA population to move from one target to the other, effectively broadcasting the perturbation from ceRNA $j$ to ceRNA $i$ through miRNA-mediated interactions.
A more direct description of this mechanism is attained instead via {\it susceptibilities} like \cite{figl1}
\begin{equation}\label{chi}
\chi_{ij}=\frac{\partial \avg{m_i}}{\partial b_j}\geq 0~~,
\end{equation} 
where $b_j$ stands for the transcription rate of ceRNA $j$. $\chi_{ij}$ quantifies the shift in the mean level of ceRNA $i$ caused by a (small) variation in $b_j$, and a large $\chi_{ij}$ (assuming no direct control of ceRNA $i$ by ceRNA $j$) points to miRNA-mediated crosstalk between ceRNAs $i$ and $j$  (see Fig. \ref{zero}C). 

While both $\chi_{ij}$ and $\rho_{ij}$ capture aspects of ceRNA crosstalk seen in experiments, their underlying physical meaning is {\it a priori} different. Fluctuating miRNA levels naturally correlate co-regulated targets, so that a large $\rho_{ij}$ is obtained when both ceRNAs respond to the stochastic dynamics of their regulator. This however does not necessarily imply a large $\chi_{ij}$. In fact, $\chi_{ij}$ can be large even in absence of fluctuations in miRNA levels, i.e. as a consequence of competition alone. In such conditions, $\rho_{ij}$ vanishes. $\chi_{ij}$ has indeed been found to be asymmetric under exchange of its indices (i.e. $\chi_{ij}\neq\chi_{ji}$ in general) \cite{figl1}, at odds with $\rho_{ij}$ which is necessarily symmetric. It would therefore be important to clarify how quantities like (\ref{rho}) and (\ref{chi}) are related in miRNA-ceRNA networks, especially to understand whether responses to perturbations (a central quantity of interest for many potential applications of the ceRNA effect) can be encoded in quantities as intuitive and as simple to measure experimentally or from data as a Pearson correlation coefficient. 

Here we show that the information conveyed by $\chi_{ij}$ is indeed captured by a correlation function similar to $\rho_{ij}$. On the other hand, $\rho_{ij}$ is linked to a susceptibility, i.e. to the response of a target to a perturbation altering the level of its competitor, but the perturbation concerns the intrinsic {\it decay} rate of the competitor rather than its transcription (as is the case for $\chi_{ij}$). In the following, we will derive these results and validate them by computer simulations and gene expression data analysis, and explore their consequences.

\section*{\textsf{Results}}

\subsection*{\textsf{Theory}}

We start from the dynamics of molecular populations in a miRNA-ceRNA network, denoting by $m_i$ the level of ceRNA species $i$ (ranging from 1 to $N$), by $\mu_a$ the level of  miRNA species $a$ (ranging from 1 to $M$), and by $c_{ia}$ the levels of miRNA-ceRNA complexes. In the deterministic limit where stochastic fluctuations are neglected, the time evolution of concentration variables is described by 
\begin{gather}
\dot \mu_a=\beta_a-\delta_a \mu_a-\sum_i k_{ia}^+ m_i\mu_a + \sum_{i}(k_{ia}^-+\kappa_{ia})c_{ia}\nonumber\\
\dot m_i=b_i-d_i m_i-\sum_a k_{ia}^+ m_i\mu_a +\sum_a k_{ia}^- c_{ia} \label{uno}\\
\dot c_{ia}= k_{ia}^+ m_i \mu_a-\frac{c_{ia}}{\tau_{ia}}\nonumber~~,
\end{gather}
with the different parameters denoting intrinsic synthesis ($b_i$, $\beta_a$) and degradation rates ($d_i$, $\delta_a$), complex association/dissociation rates ($k_{ia}^{\pm}$) and complex processing rates ($\sigma_{ia}$ and $\kappa_{ia}$ for stoichiometric and catalytic processing, respectively), while $\tau_{ia}=(\sigma_{ia}+\kappa_{ia}+k_{ia}^-)^{-1}$ represents the mean lifetime of the complex formed by miRNA $a$ and ceRNA $i$.  We note that if the mean lifetime of complexes is much shorter than that of free molecular species, i.e. if $\tau_{ia}\ll 1/d_i$ and $\tau_{ia}\ll 1/\delta_a$ for each $i$ and $a$, miRNA-ceRNA complexes achieve a steady state much faster than miRNA and ceRNA levels. In such conditions, $\dot c_{ia}\simeq 0$ and one can eliminate complexes from (\ref{uno}) by replacing $c_{ia}$ with its steady state value
\begin{equation}
\avg{c_{ia}}= k_{ia}^+ \tau_{ia} m_i \mu_a~~.
\end{equation}
For $\kappa_{ia}+k_{ia}^-\ll\sigma_{ia}$ (i.e. when stoichiometric degradation without miRNA recycling is the dominant channel of complex processing), this allows to re-cast (\ref{uno}) in the form (see Supporting Text)
\begin{equation}
\begin{aligned}
\dot m_i &\simeq -m_i\frac{\partial L}{\partial m_i}~~,\\
\dot \mu_a &\simeq -\mu_a\frac{\partial L}{\partial \mu_a}~~,\label{quattro2}
\end{aligned}
\end{equation}
where $L$ is a function of all miRNA levels $\boldsymbol{\mu}=\{\mu_a\}$ and all ceRNA levels $\mathbf{m}=\{m_i\}$ given by
\begin{multline}
L=-\sum_i (b_i\log m_i-d_i m_i)\\-\sum_a (\beta_a\log \mu_a-\delta_a \mu_a)+\sum_{i,a}k_{ia}^+ m_i\mu_a~~\label{elle}.
\end{multline}
One easily sees (see Supporting Text) that $L$ decreases along trajectories of (\ref{quattro2}), implying that its minimum describes the physically relevant steady state of (\ref{uno}) with $\mathbf{m\neq 0}$ and $\boldsymbol{\mu\neq 0}$. 

If intrinsic molecular noise (arising from stochastic transcription and degradation events and from titration due to miRNA-ceRNA interactions) is added to (\ref{uno}), after a transient molecular levels will eventually stabilize and fluctuate over time around the steady state described by the minimum of $L$. We are interested in finding a compact and intuitive mathematical form for the correlations arising between the different components in such conditions. Molecular noise is Poissonian, namely the strength of fluctuations affecting each variable is proportional to the square root of mean molecular levels (see e.g. \cite{marti} for an explicit representation in the context of a miRNA-ceRNA network), which makes our goal especially challenging. However we will see that the effects of molecular noise can be remarkably well approximated by a uniform ``effective temperature'' $T$ representing the strength of fluctuations affecting all molecular species involved. In this case, one can describe fluctuations around the steady state as thermal fluctuations around a Boltzmann-Gibbs equilibrium state. This allows to compute averages of generic functions of $\mathbf{m}$ and $\boldsymbol{\mu}$ as ``thermal averages'', i.e.
\begin{equation}
\avg{f}=\frac{1}{Z(T)}\sum_{\mathbf{m},\boldsymbol{\mu}} f(\mathbf{m},\boldsymbol{\mu}) e^{-L/T}~~,
\end{equation}
where 
\begin{equation}
Z(T)=\sum_{\mathbf{m},\boldsymbol{\mu}} e^{-L/T}
\end{equation}
is a normalization factor, the deterministic limit being obtained for $T\to 0$. In particular, defining 
\begin{equation}
\avg{fg}_c\equiv\avg{fg}-\avg{f}\avg{g}~~,
\end{equation}
by straightforward calculations one finds  
\begin{gather}
\avg{m_i}=-T\frac{\partial}{\partial d_i}\log Z(T)~~,\\
\avg{m_i m_j}_c=-T\frac{\partial\avg{m_i}}{\partial d_j}\equiv -T\omega_{ij}~~,\label{Comega}\\
\avg{m_i\log m_j}_c=T \chi_{ij}~~.
\label{chiX}
\end{gather}

Therefore, in this approximation, the susceptibility $\chi_{ij}$ [Eq. (\ref{chi})] is linked to the correlation function [Eq. (\ref{chiX})] 
\begin{equation}\label{XX}
X_{ij}=\avg{m_i\log m_j}_c 
\end{equation}
which, as $\chi_{ij}$, is not symmetric under the exchange of $i$ and $j$, while the ceRNA-ceRNA covariance 
\begin{equation}\label{CC}
C_{ij}=\avg{m_i m_j}_c
\end{equation}
is tied to the susceptibility $\omega_{ij}$ quantifying the change in $\avg{m_i}$ induced by a (small) change of the {\it intrinsic degradation rate} $d_j$ of ceRNA $j$ [Eq. (\ref{Comega})]. (Note that $\omega_{ij}\leq 0$.) The constant linking these quantities is the temperature $T$ quantifying the strength of the uniform ``effective noise''.

Somewhat unexpectedly, the above results suggest that the susceptibility $\omega_{ij}$ must be symmetric under exchange of $i$ and $j$, i.e., for instance, if the level of ceRNA $i$ is altered by changing the intrinsic degradation rate of ceRNA $j$, then the reverse is also true. To check this property, one can calculate $\omega_{ij}$ explicitly for a system formed by $N$ ceRNA species interacting with a single miRNA species at steady state by following a different route, specifically along the lines of \cite{figl1}. Considering the repression strength to which ceRNAs $i$ and $j$ are subject at a given (mean) level $\avg{\mu_1}$ of miRNA species 1 ($M=1$ in this case), one finds, for each ceRNA, a soft ``threshold'' value of $\avg{\mu_1}$, denoted by $\mu_{0,i1}\simeq d_i/k_{i1}^+$, such that $i$ is unrepressed (resp. repressed or susceptible to changes in $\avg{\mu_1}$) if $\avg{\mu_1}\ll\mu_{0,i1}$ (resp. $\gg\mu_{0,i1}$ or $\simeq \mu_{0,i1}$). A direct calculation (see Supporting Text) shows that $\omega_{ij}$ can attain large values only if both ceRNAs are susceptible to $\avg{\mu_1}$, in which case one has ($i\neq j$)
\begin{gather}
\omega_{ij}\simeq -\frac{\avg{\mu_1}}{\mu_{0,i1}\mu_{0,j1}}\frac{b_i b_j}{d_i d_j} \left(\delta_1+\sum\limits_{\ell}a_\ell\frac{b_\ell k_{\ell 1}^+}{d_\ell}\right)^{-1}~~,\label{omegaN1}
\end{gather}
where $a_\ell=0,1,1/4$ if ceRNA $\ell$ is repressed, unrepressed or susceptible, respectively. Eq. (\ref{omegaN1}) confirms that $\omega_{ij}$ is indeed symmetric under exchange of $i$ and $j$.

Concerning the approximations under which the the above results  were obtained, we remark that we started by considering (\ref{uno}) in the limit of (i) fast complex equilibration, and (ii) miRNA-ceRNA complex processing dominated by the stoichiometric channel, with the former playing the key role in deriving the function $L$ (see Supporting Text). We note however that the overall scenario just described also holds for when complexes evolve over time scales much longer than those of free molecular levels, i.e. for $\tau_{ia}\gg 1/d_i$ and $\tau_{ia}\gg 1/\delta_a$. In particular, (\ref{uno}) can again be re-cast in the form of (\ref{quattro2}) with $L$ given by (\ref{elle}), albeit with re-scaled transcription rates (see Supporting Text for details).  

Therefore we conclude that, as long as molecular noise can be approximated by a uniform effective temperature,
\begin{enumerate}
\item[(i)] the ceRNA-ceRNA covariance $C_{ij}=\avg{m_i m_j}_c$ is a proxy for the susceptibility $\omega_{ij}$, and\\
\item[(ii)] the correlation function $X_{ij}=\avg{m_i \log m_j}_c$ is a proxy for the susceptibility $\chi_{ij}$. 
\end{enumerate}

\subsection*{\textsf{Validation}}

\begin{figure*}
\begin{center}
\includegraphics[width=0.8\textwidth]{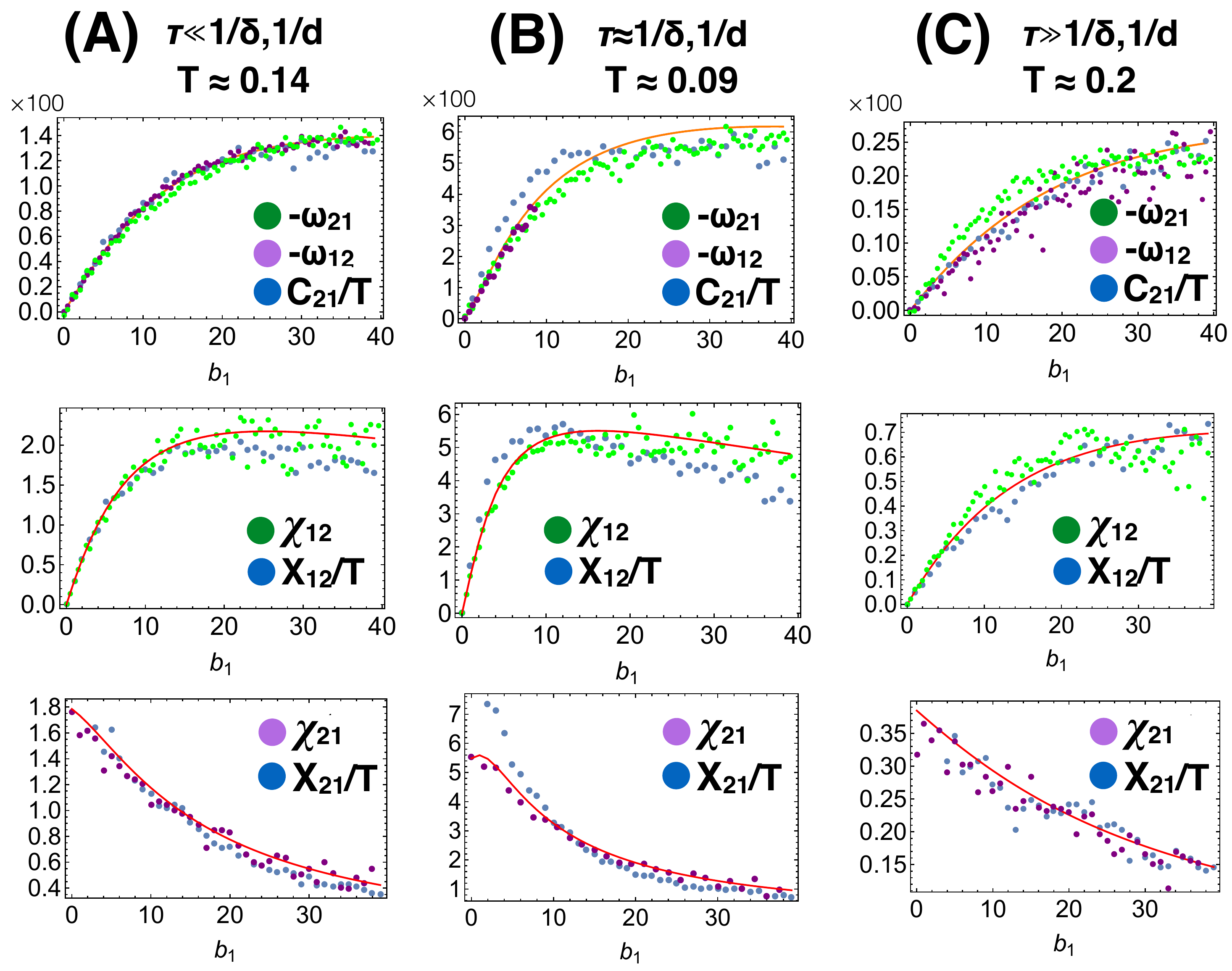}
\end{center}
\caption{Results from Gillespie simulations (markers) versus analytical predictions for susceptibilities (lines) for a miRNA-ceRNA network with $N=2$ ceRNA and $M=1$ miRNA species and different values of the transcription rate of ceRNA species 1, $b_1$. Numerical susceptibilities have been obtained by perturbing the system at steady state and recording the new steady state at which the system settles after a transient. Analytical lines correspond to the numerical derivatives of the steady state levels obtained from (\ref{uno}). Sets (A)--(C) are for different values of the mean lifetime of miRNA-ceRNA complexes $\tau$, and refer, respectively, to the cases in which complex processing is much faster, comparable or much slower, than the degradation dynamics of free miRNAs and ceRNAs. Also reported is the value of $T$ that provides the best fit in each case. See Table I for parameter values. Results obtained for stronger miRNA-ceRNA couplings are shown in the Supporting Text. \label{figuno}}
\end{figure*}

We have validated the above scenario by simulating a small network involving 2 ceRNA and a single miRNA species via the Gillespie algorithm \cite{gill}, where molecular noise is accounted for explicitly (see Supporting Text). Results are summarized in Fig. \ref{figuno}, where we compare $\omega_{12}$, $\omega_{21}$ and $C_{12}\equiv C_{21}$ on on hand, and $\chi_{12}$, $\chi_{21}$, $X_{12}$ and $X_{21}$ on the other, as computed from simulations (i.e. with the actual molecular noise), against the theoretical predictions. We considered three scenarios for the mean lifetime of miRNA-ceRNA complexes, namely those covered by the theory (i.e. complex equilibration much faster and much slower than the equilibration of miRNA and ceRNA levels) as well as the intermediate case where characteristic timescales are comparable for all variables.

\begin{table}[b]
\begin{tabular}{ c c c c c }
\rowcolor{gray!20}
Parameters [units] & Figs 2A & Figs 2B &  Figs 2C\\
\hline
\hline
$b_2$, $\beta_1$ [molec. min$^{-1}$]	 & 10 & 10& 10	    \\ 
\rowcolor{gray!15}
$k_{11}^+$	[molec.$^{-1}$ min$^{-1}$]  	&	$e^{-5}$ &	$e^{-5}$  &	$e^{-5}$  \\
$k_{21}^+$	[molec.$^{-1}$ min$^{-1}$] 	&	$e^{-6}$	&	$e^{-6}$ &		$e^{-6}$ \\  
\rowcolor{gray!15}
$k_{11}^-$, $k_{21}^-$	[min$^{-1}$]  	&	0.001 &	0.001	&	0.001 \\
$d_1$, $d_2$, $\delta_1$ [min$^{-1}$]		& 	0.1 &	0.05	&	0.2	 \\
\rowcolor{gray!15}
$\sigma_{11}$, $\sigma_{21}$ [min$^{-1}$] 	&	1	&	0.05	&	0.05 	\\
$\kappa_{11}$, $\kappa_{21}$ [min$^{-1}$]  	&	0.001 &	0.001	&	0.001		\\
\hline
\end{tabular}
\caption{Parameters used in Fig \ref{figuno}. Note that $i\in\{1,2\}$ while $a=1$. \label{Tab:table1}}
\end{table}

One sees that theoretical predictions obtained in the ``thermal noise'' approximation agree remarkably well with simulations including the actual molecular noise. In particular, the full correspondence between the susceptibilities $\omega_{ij}$ and $\chi_{ij}$ and the (re-scaled) correlation functions $C_{ij}$ and, respectively, $X_{ij}$ is evident. Notice that a single global parameter $T\geq 0$ has been used to fit all data in each of the conditions. This shows how accurately the assumption of a uniform effective temperature can mimic the effects of intrinsic stochasticity. On the other hand, its limits might be reflected, at least in part, in the discrepancies that occur at high transcription rates.

These results confirm that (\ref{chiX}) and (\ref{CC}) are indeed good predictors of the response of a ceRNA to a perturbation affecting one of its competitors within a miRNA-ceRNA network. Notably, such correlation functions are easy to estimate from transcription data sets. Our framework therefore has the potential to offer new insight into post-transcriptional regulation, its system-level organization and its impact on cellular functions. 

\begin{figure*}
\begin{center}
\includegraphics[width=0.99\textwidth]{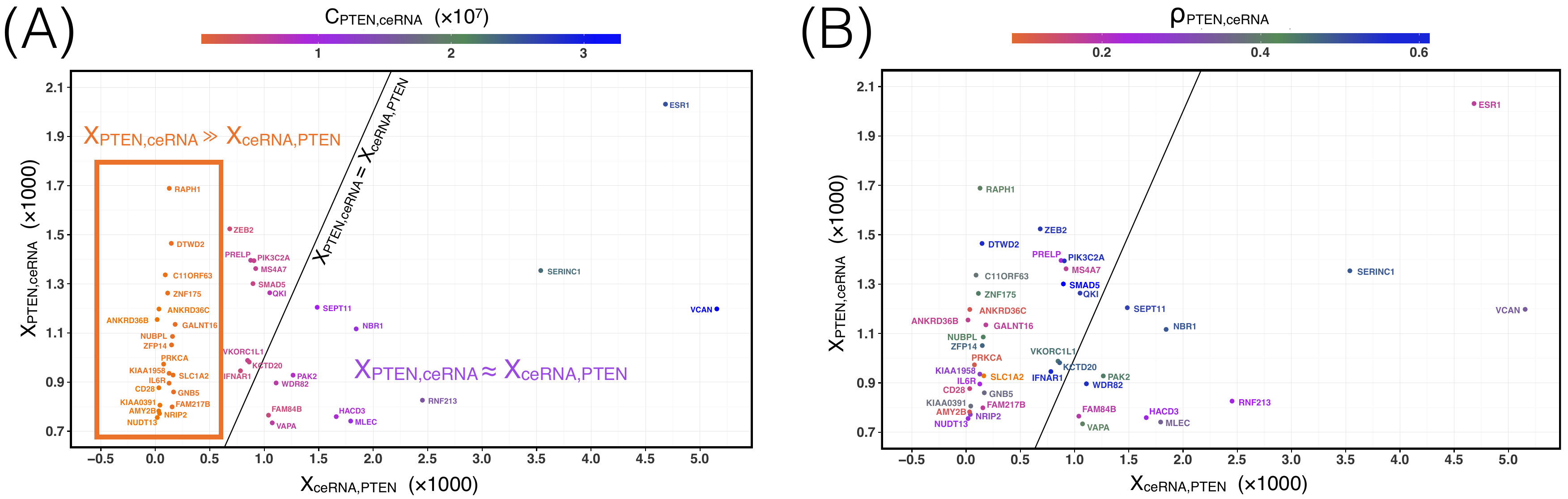}
\end{center}
\caption{Values of different correlation functions of PTEN with its ceRNAs computed from breast cancer samples from TCGA. (A) $X_{\text{ceRNA,PTEN}}$ and $X_{\text{PTEN,ceRNA}}$ are reported on the $x$ and $y$ axes respectively, while the color code gives the value of $C_{\text{PTEN,ceRNA}}$. (B) Same as (A), except that the color code now corresponds to the Pearson coefficient $\rho_{\text{PTEN,ceRNA}}=C_{ij}/(\sigma_{m_i}\sigma_{m_j})$. Notice that in (B) PTEN-ceRNA pairs occupy the same positions in the plane as in (A). See Supporting Text for higher-resolution versions of both panels. \label{figtre}}
\end{figure*}

In order to test this idea, we analyzed the ceRNA scenario emerging from 1098 breast cancer samples obtained from The Cancer Genome Atlas \cite{tcga}, focusing on the widely studied oncosuppressor PTEN and its immediate competitors (i.e. the ceRNAs sharing at least one miRNA regulator with PTEN). In particular, we computed  
\begin{gather}
C_{\text{PTEN,ceRNA}}=\avg{m_\text{PTEN} m_\text{ceRNA}}_c \label{cij}\\
X_{\text{ceRNA,PTEN}}=\avg{m_\text{ceRNA} \log m_\text{PTEN}}_c\label{xcp}\\
X_{\text{PTEN,ceRNA}}=\avg{m_\text{PTEN} \log m_\text{ceRNA}}_c\label{xpc}
\end{gather}
for a set of candidate PTEN ceRNAs found in \cite{tay} by means of Mutually Targeted miRNA-Response Element Enrichment Analysis. Notice that the average appearing in Eqs. (\ref{cij}--\ref{xpc}) is over samples and not over time. We expect however that, if the interaction network is conserved across samples, averages over samples should reproduce statistical averages such as (\ref{XX}),  as different samples effectively represent different snapshots of the state of the network. Fig. \ref{figtre}A shows that when (\ref{cij}) (whose value is encoded in the color of markers) is large, both (\ref{xcp}) and (\ref{xpc}) tend to be large. According to (\ref{omegaN1}), a large $C_{ij}$ (or $\omega_{ij}$) signals that both PTEN and its competitor are susceptible to changes in the level of at least one of their shared regulators.
For such pairs, in addition, it has previously been shown that both $\chi_{ij}$ and $\chi_{ji}$ are expected to be large \cite{figl1}. This implies a fully bi-directional crosstalk, i.e. any perturbation affecting the level of one species should affect the level of the other via miRNA-mediated regulation. Remarkably, this was experimentally shown to be the case in \cite{tay} for some of the ceRNAs we tested (e.g. SERINC1, VAPA), all of which are in this regime according to our analysis. Adding to this, we are also able to point to a number of other PTEN competitors, a perturbation of which should trigger a response by PTEN.

On the other hand, smaller values of (\ref{cij}) (orange markers in Fig. \ref{figtre}A) are associated to strongly asymmetric PTEN-ceRNA pairs for which (\ref{xpc}) is much larger than (\ref{xcp}). This suggests that PTEN will respond to an increase of its competitor's bare transcription rate (and not vice-versa), while no response of PTEN should be expected upon perturbing   the bare decay rate of the same ceRNA as $C_{ij}$ is small. Within the steady state theory of \cite{figl1}, ceRNA pairs with strongly different values of $\chi_{ij}$ and $\chi_{ji}$ pertain to cases where the responding ceRNA (PTEN here) is susceptible to variations in the miRNA levels while the perturbed one (PTEN's competitor) is fully repressed. Our data analysis fully confirms both this scenario and the theory presented here in linking such cases to low values of the bare covariance (\ref{CC}).

Finally, note (Fig. \ref{figtre}B) that the above information can  not be retrieved if $C_{ij}$ is replaced by the Pearson coefficient  $\rho_{ij}$, Eq. (\ref{rho}), which just amounts to normalizing $C_{ij}$ by the product of the standard deviations $\sigma_{m_i}$ and $\sigma_{m_j}$ of $m_i$ and $m_j$. Indeed, using the value of $\rho_{ij}$ to color-code PTEN's ceRNAs, one sees that the Pearson coefficient can mislead into expecting (or not expecting) a response to a perturbation when the actual susceptibilities are small (resp. large). 

For instance, $\rho_{ij}$ is rather small for the pair formed by PTEN and SLC1A2, which seems to suggest absence of mutual cross-talk between these two transcripts. However, while both $C_{\text{PTEN,SLC1A2}}$ and $X_{\text{SLC1A2,PTEN}}$ are small, $X_{\text{PTEN,SLC1A2}}$ is significant. This suggests  that (i) SLC1A2 will not respond to a perturbation affecting the transcription rate of PTEN, and (ii) the pair will be insensitive to changes in each other's bare decay rate; however, (iii) PTEN {\it will be affected} by a change in the bare transcription rate of its competitor despite the small statistical correlation that exists between their levels. Likewise, the large value of the Pearson coefficient between PTEN and DTWD2 can mislead into generically expecting a response when instead the susceptibility is strongly perturbation-dependent. In particular, the level of DTWD2 should not be significantly modified by a change in the level of PTEN (as $X_{{\rm DTWD2,PTEN}}$ is rather small) in spite of the large Pearson coefficient. Notice that, remarkably, for this pair, $C_{ij}$ and $\rho_{ij}$ take on very different values.

\section*{\textsf{Discussion}}

To sum up, we have identified [Eq.s (\ref{Comega}) and (\ref{chiX})] a set of correlation functions that can serve as proxies for ceRNA susceptibilities to perturbations. Specifically, $C_{ij}=\avg{m_i m_j}_c$ is related to the susceptibility $\omega_{ij}$ quantifying ceRNA $i$'s response to a change of the bare decay rate of ceRNA $j$, while  $X_{ij}=\avg{m_i\log m_j}_c$ is related to the susceptibility $\chi_{ij}$ quantifying ceRNA $i$'s response to a change of the bare transcription rate of ceRNA $j$. These relations are valid at steady state and within the approximations discussed, are fully confirmed by numerical simulations. 

Most importantly, quantities like $C_{ij}$ and $X_{ij}$ can be easily estimated from data and possibly measured in experiments. An analysis of PTEN's emergent crosstalk pattern from TCGA breast cancer dataset using these functions has indeed shown that a map of ceRNA responses to perturbations affecting competitors can be constructed by combining the information provided by each, while the Pearson coefficient $\rho_{ij}$ can be inaccurate in this respect. This opens the way to probing the structure and function of ceRNA networks {\it in silico} by straightforwardly analyzing transcriptional data, and provides a key to obtain testable transcriptome-scale predictions about ceRNA crosstalk. 

Notice that our results apply without any modification to ceRNA pairs that don't share miRNA regulators, i.e. it is capable of identifying long-range crosstalk (i.e. interactions between ceRNAs that are separated by multiple miRNAs along the miRNA-ceRNA network) of the kind discussed in \cite{biham}. 

From the viewpoint of physics, results like (\ref{Comega}) and (\ref{chiX}) are akin to the ``fluctuation-response relations'' that constitute a cornerstone of statistical mechanics \cite{betto}. Their derivation in our context has relied on an equilibrium framework that presupposes stationarity of molecular levels. Since ceRNA crosstalk can be substantially more complex away from the steady state \cite{figl2}, a more refined mathematical study will be required to extend the theory developed here to off-equilibrium dynamical regimes. Our results on the other hand may also open the way to the application of  recently developed inference techniques \cite{inf} to estimate miRNA levels or kinetic parameters from ceRNA levels. 

\subsection*{\textsf{Acknowledgments}}

We gratefully acknowledge Carla Bosia and Andrea Pagnani for useful insight and suggestions.

\newpage

\widetext

\centerline{\textbf{SUPPORTING TEXT}}

\section{Derivation of the function $L$ for fast complex processing}

Starting from Eq. (3) of the Main Text, namely (with $\tau_{ia}=(\sigma_{ia}+\kappa_{ia}+k_{ia}^-)^{-1}$)
\begin{gather}
\dot \mu_a=\beta_a-\delta_a \mu_a-\sum_i k_{ia}^+ m_i\mu_a + \sum_{i}(k_{ia}^-+\kappa_{ia})c_{ia}~~,\nonumber\\
\dot m_i=b_i-d_i m_i-\sum_a k_{ia}^+ m_i\mu_a +\sum_a k_{ia}^- c_{ia}~~, \label{unoaa}\\
\dot c_{ia}= k_{ia}^+ m_i \mu_a-\frac{c_{ia}}{\tau_{ia}}\nonumber~~,
\end{gather}
we assume that complexes equilibrate much faster than miRNA and ceRNA levels and substitute $c_{ia}$ with its steady state value $\avg{c_{ia}}= k_{ia}^+ \tau_{ia} m_i \mu_a$. One finds
\begin{equation}
\begin{aligned}
\dot \mu_a\simeq\beta_a-\delta_a \mu_a-\sum_i \frac{\sigma_{ia}}{\sigma_{ia}+\kappa_{ia}+k_{ia}^-} \,\,k_{ia}^+ m_i\mu_a ~~,\\
\dot m_i\simeq b_i-d_i m_i-\sum_a \frac{\sigma_{ia}+\kappa_{ia}}{\sigma_{ia}+\kappa_{ia}+k_{ia}^-}\,\,k_{ia}^+ m_i\mu_a~~.
\end{aligned}
\end{equation}
For $\kappa_{ia}+k_{ia}^-\ll\sigma_{ia}$, this reduces to
\begin{equation}
\begin{aligned}
\dot \mu_a\simeq\beta_a-\delta_a \mu_a-\sum_i k_{ia}^+ m_i\mu_a~~, \\
\dot m_i\simeq b_i-d_i m_i-\sum_a k_{ia}^+ m_i\mu_a~~,
\end{aligned}\label{fort}
\end{equation}
which is easily seen to be equivalent to Eq. (5) in the Main Text, with
\begin{equation}
L=-\sum_i (b_i\log m_i-d_i m_i)-\sum_a (\beta_a\log \mu_a-\delta_a \mu_a)+\sum_{i,a}k_{ia}^+ m_i\mu_a~~\label{elle}.
\end{equation}

\section{$L$ decreases along the dynamics}

By direct differentiation and using the fact that (see Eq. (5) in the Main Text)
\begin{equation}
\begin{aligned}
\frac{\dot m_i}{m_i} &\simeq -\frac{\partial L}{\partial m_i}~~,\\
\frac{\dot \mu_a}{\mu_a} &\simeq -\frac{\partial L}{\partial \mu_a}~~,\label{quattro2}
\end{aligned}
\end{equation}
one finds
\begin{equation}
\dot{L}=\sum_i\frac{\partial L}{\partial m_i}\dot{m_i}+\sum_a\frac{\partial L}{\partial\mu_a}\dot{\mu_a}=-\sum_i\frac{\dot{m_i}^2}{m_i}-\sum_a\frac{\dot{\mu_a}^2}{\mu_a}\leq 0~~.
\end{equation}
In other words, under the approximations discussed above, $L$ decreases along the dynamics of the miRNA-ceRNA network. Therefore the minimum of $L$ (which is unique by virtue of the concavity of $L$) describes a steady state of the dynamics (\ref{unoaa}).

\section{Approximate calculation of $\omega_{ij}$ for a system with one miRNA  and $N$ ceRNA species}

Starting from Eq. (\ref{fort}) taken for $N$ ceRNA species and a single miRNA species (we suppress its index for simplicity) in the limit $\sigma_i\gg k_i^-+\kappa_i$, the steady-state level of $m_i$ reads
\begin{gather}
m_i=\frac{b_i}{d_i}F_i(\mu)~~~~~,~~~~~ F_i(\mu)=\frac{\mu_{0,i}}{\mu_{0,i}+\mu}~~~~~,~~~~~\mu_{0,i}=\frac{d_i}{k_i^+}~~.
\end{gather}
Now noting that
\begin{gather}
F_i(\mu)\simeq
\begin{cases}
1-\frac{\mu}{\mu_{0,i}}&\text{for }\mu\ll\mu_{0,i}~~\text{(ceRNA $i$ expressed)}\\
\frac{1}{2}-\frac{\mu-\mu_{0,i}}{4\mu_{0,i}}&\text{for }\mu\simeq\mu_{0,i}~~\text{(ceRNA $i$ susceptible)}\\
\frac{\mu_{0,i}}{\mu}&\text{for }\mu\gg\mu_{0,i}~~\text{(ceRNA $i$ repressed)}
\end{cases}\\
\frac{dF_i}{d\mu}\equiv F_i'\simeq
\begin{cases}
-\frac{1}{\mu_{0,i}}&\text{for }\mu\ll\mu_{0,i}~~\text{(ceRNA $i$ expressed)}\\
-\frac{1}{4\mu_{0,i}}&\text{for }\mu\simeq\mu_{0,i}~~\text{(ceRNA $i$ susceptible)}\\
-\frac{\mu_{0,i}}{\mu^2}&\text{for }\mu\gg\mu_{0,i}~~\text{(ceRNA $i$ repressed)}
\end{cases}
\end{gather}
and that the steady state miRNA level can be approximated by \cite{figl1}
\begin{gather}
\mu\simeq\frac{\beta-\sum_{i\in\text{Repr}}b_i-\frac{1}{4}\sum_{i\in\text{Susc}}b_i}{\delta+\sum_{i\in\text{Expr}}\frac{b_i k_i^+}{d_i}+\frac{1}{4}\sum_{i\in\text{Susc}}\frac{b_i k_i^+}{d_i}}~~,
\end{gather}
so that
\begin{gather}
\frac{\partial\mu}{\partial d_j}\simeq
\chi_{\mu\mu}\frac{b_j}{d_j}\frac{\mu}{\mu_{0,j}}\times
\begin{cases}
1 &\text{for }\mu\gg\mu_{0,i}~~\text{(ceRNA $i$ expressed)}\\
\frac{1}{4}&\text{for }\mu\simeq\mu_{0,i}~~\text{(ceRNA $i$ susceptible)}\\
0&\text{for }\mu\gg\mu_{0,i}~~\text{(ceRNA $i$ repressed)}
\end{cases}\\
\chi_{\mu\mu}=\left(\delta+\sum_{i\in\text{Expr}}\frac{b_i k_i^+}{d_i}+\frac{1}{4}\sum_{i\in\text{Susc}}\frac{b_i k_i^+}{d_i}\right)^{-1}~~,
\end{gather}
we can compute the susceptibility $\omega_{ij}$ as
\begin{gather}
\omega_{ij}\equiv\frac{\partial m_i}{\partial d_j}=
\frac{b_i}{d_i}F_i'\frac{\partial\mu}{\partial d_j}~~~~~~~~~~~(i\neq j)~~ .
\end{gather}
One finds
\begin{gather}
\omega_{ij}\simeq
-\chi_{\mu\mu}\frac{b_i b_j}{d_i d_j}W_{R(i),R(j)}~~~~~~~~~~~(i\neq j)~~ ,
\end{gather}
where $\hat{W}$ is a $3\times 3$ matrix that only depends on the regime $R(i)$ (repressed, susceptible or expressed) to which ceRNA $i$ belongs. By considering the definitions of the different regimes in terms of the value of $\mu$, all elements of $\hat{W}$ are found to be $\ll 1$ (for instance, $W_{{\rm Expr,Expr}}=\mu/(\mu_{0,i}\mu_{0,j})\ll 1$ as $\mu\ll\mu_{0,i}$ and $\mu\ll\mu_{0,j}$ if ceRNAs $i$ and $j$ are both expressed) except for $W_{{\rm Susc,Susc}}$, which is given by
\begin{gather}
W_{{\rm Susc,Susc}}=\frac{1}{16}\frac{\mu}{\mu_{0,i}\mu_{0,j}}~~,
\end{gather}
leading immediately to Eq. (15) of the Main Text. 

\section{Case of slow complex processing}

Assuming complex levels $c_{ia}$ are roughly stationary over time scales for which $m_i$ and $\mu_a$ evolve (i.e. $\tau_{ia}\gg 1/d_i$ and $\tau_{ia}\gg 1/\delta_a$ for each $i$ and $a$), then all terms in (\ref{unoaa}) that involve the variables $c_{ia}$ can be taken to be roughly constant for short enough characteristic times. In such conditions, miRNAs are effectively transcribed at rates 
\begin{equation}
\beta_a^{\eff}\simeq \beta_a+\sum_{i}(k_{ia}^-+\kappa_{ia})c_{ia}~~,
\end{equation}
while ceRNAs are effectively transcribed at rates 
\begin{equation}
b_i^{\eff}\simeq b_i+\sum_a k_{ia}^- c_{ia}~~.
\end{equation}
In this limit, (\ref{unoaa}) can again be cast as
\begin{equation}
\begin{aligned}
\dot m_i\simeq -m_i\frac{\partial L}{\partial m_i}~~,\\
\dot \mu_a\simeq -\mu_a\frac{\partial L}{\partial \mu_a}~~,\label{quattro3}
\end{aligned}
\end{equation}
with
\begin{equation}\label{H}
L=-\sum_i (b_i^{\eff}\log m_i-d_i m_i)-\sum_a (\beta_a^{\eff}\log \mu_a-\delta_a \mu_a)+\sum_{i,a}k_{ia}^+ m_i\mu_a~~.
\end{equation}
The main difference from the previous case lies in the fact that the minimum of $L$ should now be computed self-consistently from the asymptotic value of $c_{ia}$: after the (fast) equilibration of $m_i$'s and $\mu_a$'s following (\ref{quattro3}), a new steady state value for complexes is computed as $c_{ia}=k_{ia}^+\tau_{ia}m_i\mu_a$, leading in turn to new values for the effective transcription rates $\beta_a^{\eff}$ and $b_i^{\eff}$ and hence to new values for $m_i$'s and $\mu_a$'s from (\ref{quattro3}), and so on until convergence.

\section{Stochastic dynamics of a miRNA-ceRNA network}

The time evolution of our miRNA-ceRNA network with $N$ ceRNA species (labeled $i$), $M$ miRNA species (labeled $a$) and intrinsic (molecular) noise is described by the system 
\begin{gather}
\dot \mu_a=\beta_a-\delta_a \mu_a-\sum_i k_{ia}^+ m_i\mu_a + \sum_{i}(k_{ia}^-+\kappa_{ia})c_{ia}+\eta_a~~,\nonumber\\
\dot m_i=b_i-d_i m_i-\sum_a k_{ia}^+ m_i\mu_a +\sum_a k_{ia}^- c_{ia}+\xi_i ~~, \label{unob}\\
\dot c_{ia}= k_{ia}^+ m_i \mu_a-\frac{c_{ia}}{\tau_{ia}}+\zeta_{ia}\nonumber~~,
\end{gather}
where $\tau_{ia}=(\sigma_{ia}+\kappa_{ia}+k_{ia}^-)^{-1}$ while $\eta_a$, $\xi_a$ and $\zeta_{ia}$ represent stochastic variables. As each noise source contributes independently to the overall noise level, one has
\begin{gather}
\eta_a=\eta_{\mu_a}- \sum_i \zeta^+_{ia} + \sum_i \zeta^-_{ia} + \sum_i \zeta^\kappa_{ia}~~,\\
\xi_i=\xi_{m_i} - \sum_a\zeta^+_{ia} + \sum_a\zeta_{ia}^-~~,\\
\zeta_{ia}=\zeta_{ia}^{\sigma} + \zeta^+_{ia} - \zeta^-_{ia} - \zeta^\kappa_{ia}
\end{gather}
where $\xi_{m_i}$, $\eta_{\mu_a}$, $\zeta_{ia}^\pm$, $\zeta_{ia}^\sigma$, and $\zeta_{ia}^\kappa$ and are mutually independent zero-average random variables representing, respectively, the intrinsic noise in ceRNA levels, in miRNA levels, in the binding/unbinding dynamics of complexes, in the stoichiometric complex degradation channel and in the catalytic complex degradation channel. Correlations are, for each component, described by
\begin{equation}
\begin{aligned}
\avg{\xi_{m_i}(t)\xi_{m_i}(t')} &=  (d_i \bar{m}_i + b_i) ~ \delta(t-t')~~,\\
\avg{\xi_{\mu_a}(t)\xi_{\mu_a}(t')} &= (\delta_a \bar{\mu}_a + \beta_a ) ~ \delta(t-t')~~, \\
\avg{\zeta_{ia}^+(t)\zeta_{ia}^+(t')} &=  k_{ia}^+ \bar{ m}_i \bar{\mu}_a ~ \delta(t-t')~~,\\
\avg{\zeta_{ia}^-(t)\zeta_{ia}^-(t')} &=  k_{ia}^- \bar{ c}_{ia} ~ \delta(t-t')~~,\\
\avg{\zeta_{ia}^\sigma(t)\zeta_{ia}^\sigma(t')} &= \sigma_{ia} \bar{c}_{ia} ~ \delta(t-t')~~,\\
\avg{\zeta_{ia}^\kappa(t)\zeta_{ia}^\kappa(t')} &=  \kappa_{ia} \bar{ c}_{ia} ~ \delta(t-t')~~,
\end{aligned}
\end{equation}
where
\begin{gather}
\label{eq:steadystateM}
\bar{m}_i = \frac{b_i  + \sum_a k_{ia}^- \bar{c}_{ia} }{d_i + \sum_a k_{ia}^+ \bar{\mu}_a }~~, \\ 
\bar{\mu}_a = \frac{\beta_a + \sum_i (k_{ia}^- + \kappa_{ia}) \bar{c}_{ia} }{ \delta_a + \sum_i k^+_{ia} \bar{m}_i }~~,\\
\bar{c}_{ia} = \frac{k^+_{ia} \bar{\mu}_a \,\, \bar{m}_i }{\sigma_{ia} + k^-_{ia} + \kappa_{ia}}~~
\label{eq:steadystateMEND}
\end{gather}
denote the mean steady-state molecular levels. To obtain Fig. 2 of the Main Text, we have simulated the above system with $M=1$, $N=2$ using the Gillespie algorithm \cite{gill}.

\vspace{3cm}

\begin{figure}[h!]
\begin{center}
\includegraphics[width=0.99\textwidth]{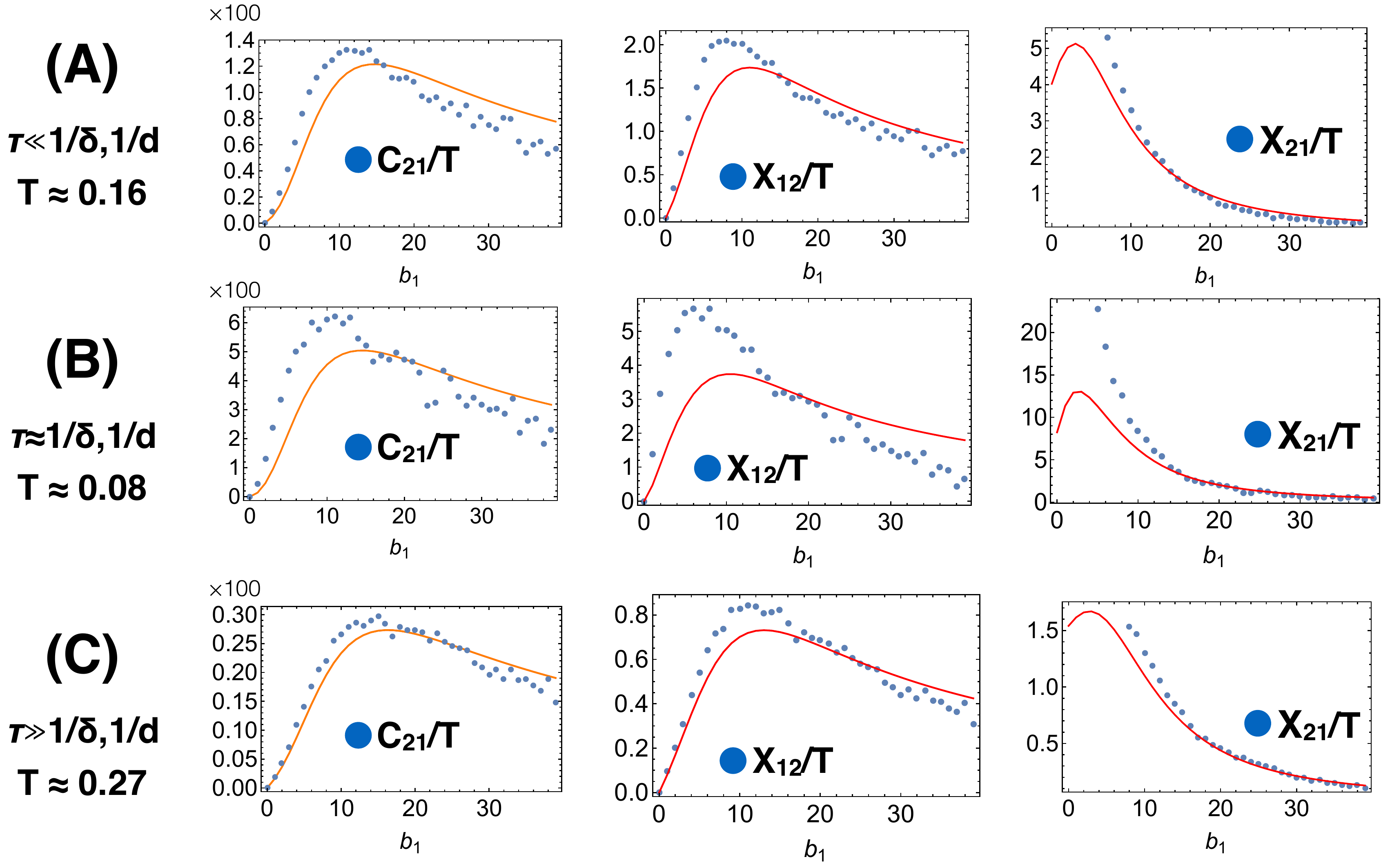}
\end{center}
\caption{Same as Figure 2 in the Main Text but for stronger miRNA repression. Parameter values are as in Table I of the Main Text except for $k_{11}^+=e^{-3}$ and $k_{21}^+=e^{-4}$. Note that, for small values of $b_1$, $m_1$ gets too small to accurately estimate $X_{ij}$s. \label{figdue}}
\end{figure}

\newpage

\begin{figure*}
\begin{center}
\includegraphics[angle=-90,width=0.8\textwidth]{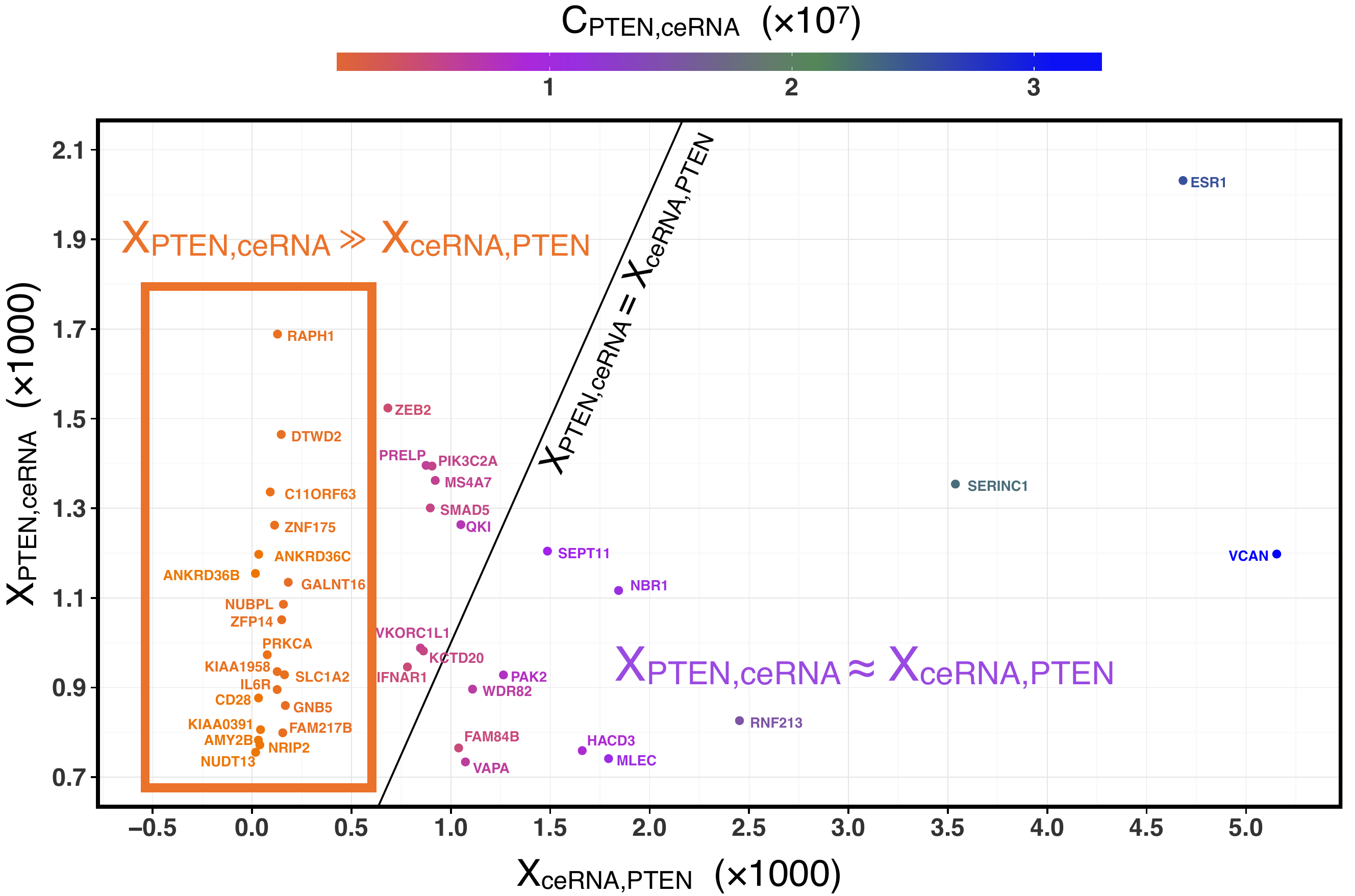}
\end{center}
\caption{Large-scale version of Fig. 3A in the Main Text.}
\end{figure*}

\newpage

\begin{figure*}
\begin{center}
\includegraphics[angle=-90,width=0.8\textwidth]{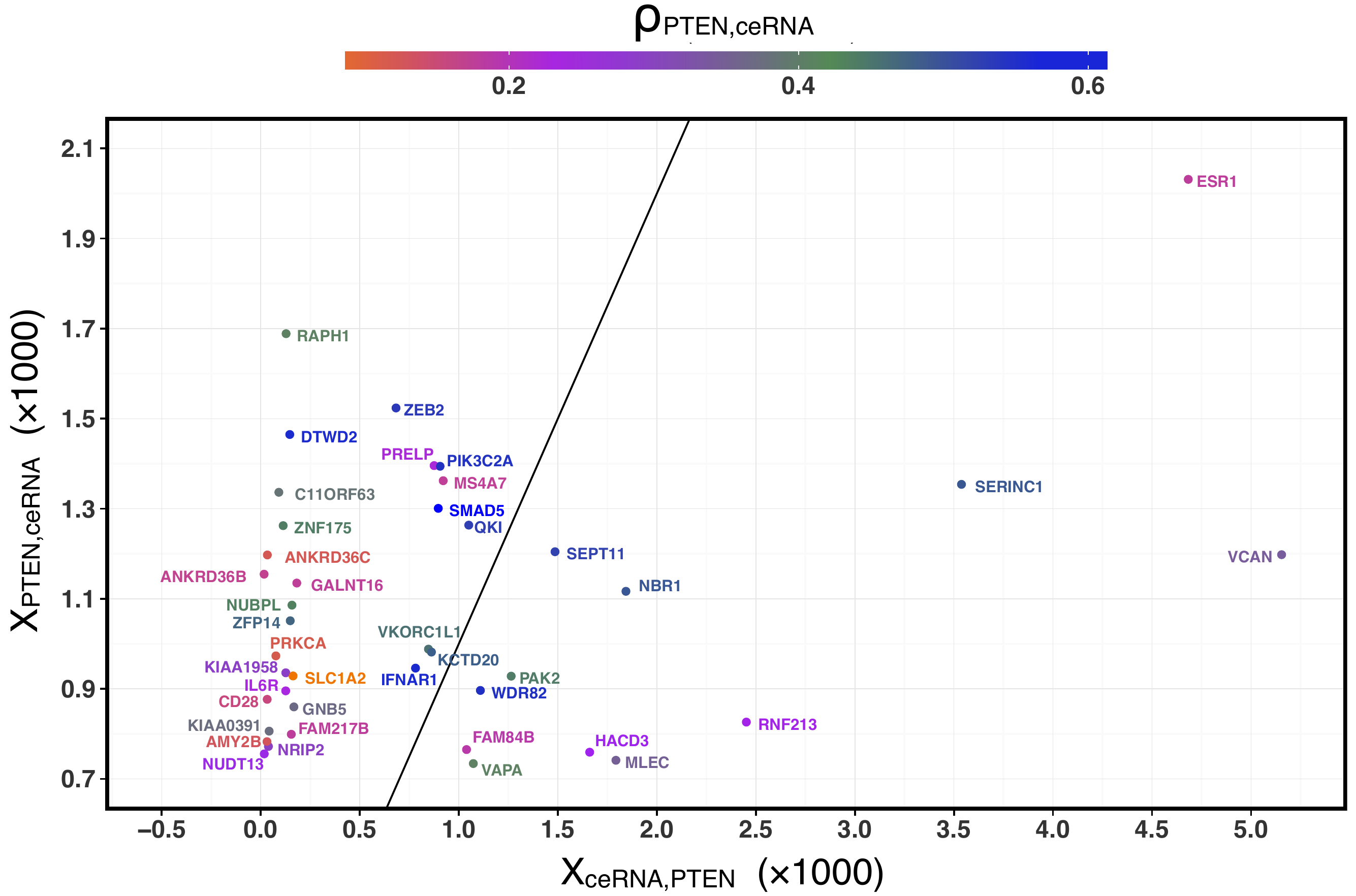}
\end{center}
\caption{Large-scale version of Fig. 3B in the Main Text.}
\end{figure*}

\end{document}